# Oscillatory Electron-Phonon Coupling in Ultra-Thin Silver Films on V(100)

T. Valla[1*], M. Kralj[2], A. Šiber[2], M. Milun[2], P. Pervan[2],
P.D. Johnson[1] and D P. Woodruff[3]

[1]*Physics Department, Brookhaven National Laboratory, Upton, NY, USA*
[2]*Institute of Physics, Zagreb, Croatia*
[3]*Physics Department, University of Warwick, Coventry, UK*

The temperature dependence of peak widths in high resolution angle-resolved photoelectron spectroscopy from quantum well states in ultra thin Ag films on V(100) has been used to determine the electron-phonon coupling constant, $\lambda$, for films of thickness 1 – 8 layers. A strong oscillatory variation in coupling strength is observed as a function of film thickness, peaking at a 2 layer film for which $\lambda \approx 1.0$. A simple theory incorporating interaction of the photo-hole with the thermal vibrations of the potential step at the adlayer-vacuum interface is shown to reproduce the main features of these results.

The physical properties of metal surfaces and thin metallic films are often different from those of the bulk. In general, reduced dimensionality and a higher degree of localisation enhance correlation in electronic systems. In this regard quantum-well (QW) states in thin films provide a testing ground of such effects, because the degree of localisation can be varied by changing the film thickness. While a significant body of previous studies have explored the one-electron behaviour of quantum-well states, there have been no comparable studies on the influence of the changing degree of localisation in ultra-thin quantum wells on the many-electron properties. Here we present data which provide a clear manifestation of essentially new phenomena in the electron-phonon coupling in such systems.

Specifically, we report measurements of the electron-phonon coupling in ultra-thin silver films deposited on a V(100) surface obtained from the temperature dependence of the widths of ARPES (angle-resolved photoelectron spectroscopy) peaks from QW states in the films. An exceptionally strong oscillatory change in the coupling strength is observed, as a function of the thickness of the silver film, with a pronounced maximum associated with the QW state corresponding to a 2 ML (monolayer) film. We show that this observed variation of the coupling strength can be explained in terms of the interaction of the photo-hole with the thermal vibration of the potential step at the adlayer-vacuum interface. The oscillatory behaviour reflects mainly the variable degree of localisation of the QW states which, through the amplitude of the wave-function at the surface-vacuum interface, influences the coupling to the surface vibrational mode. The increased electron-phonon coupling strength of these films, which at certain thicknesses substantially exceeds the value for bulk Ag, could have important consequences for the possibility of tailoring ultra-thin structures to have particular characteristics. For example, an increased superconducting transition temperature is one possibility, experimentally observed in number of systems [1], although strong electron-phonon coupling itself is not sufficient to generate a high superconducting transition temperature in thin film and surface states [2].

ARPES has already been shown to be particularly suitable for studying many-body interactions in quasi-two dimensional systems, because in such systems it measures directly the photo-hole spectral function, and there have been several such studies of metallic surface states or in thin films. A study of the Mo(110) surface state [3] showed that all three interaction terms, electron – electron, electron – phonon and defect scattering, can be deduced from the temperature and binding energy dependence of the photoemission peak width. Other studies on metallic surface states on Cu [4], Be [2, 5] and Ga [6] focused on the electron-phonon coupling term alone. The electron-phonon coupling constants determined in this way for surface states can be significantly different from those measured in the bulk. For example, the value for the Be(0001) surface state is 4 times that found in the bulk by transport measurements. ARPES has also been used to study the electron-phonon coupling in QW states in films sufficiently thick for the results to be judged to be characteristic of the bulk metal [7, 8]; in this case some differences in the values deduced have been attributed to the momentum-resolved character of an ARPES measurement as opposed to a directionally-averaged transport measurement. There has been just one previous study of electron-phonon coupling in a thin quantum well (of Na on Cu(111)) which we will discuss in the context of our own results [9].

---

[*] Corresponding author: valla@bnl.gov

The experiments reported here were carried out at the National Synchrotron Light Source using undulator beamline U13UB which provides photon energies in the range between 12 and 23 eV. The electron energy analyser was a Scienta SES-200 which collects simultaneously a large energy and angular window (~12°) of the photoelectrons. This reduces the time needed for data acquisition and ensures that a wide range of the states in k-space are recorded under identical conditions of temperature and surface cleanliness. The combined instrumental energy resolution could be set to a value in the range 8-25 meV, small enough to make no significant contribution to the measured photoemission peak width from the QW states. The angular resolution was ~0.2°. The base pressure in the experimental chamber was $4 \times 10^{-9}$ Pa. Detailed characterisation of the epitaxial growth of silver on V(100) has been presented in Refs.[10,11] while Refs.[12,13,14,15] describe in detail the QW state properties in this system. Films of 1 ML and 2 ML are stable up to 800 K [11], while thicker films are stable only up to 350 K.

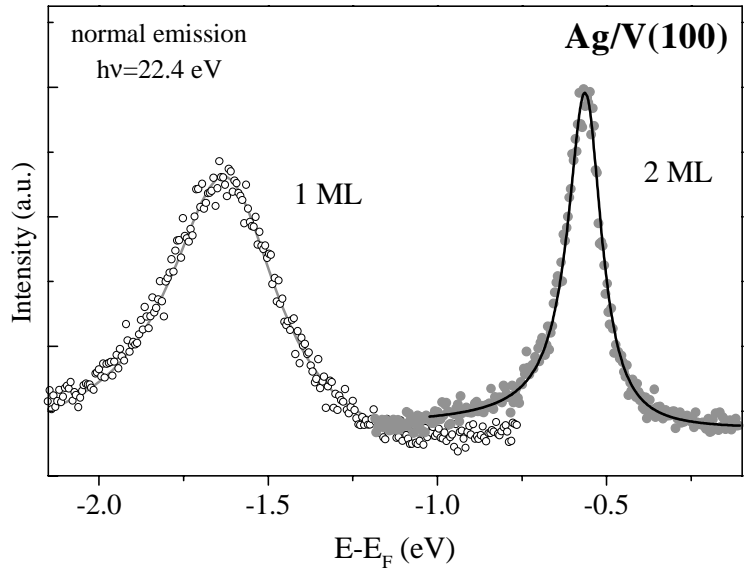

**Fig.1** Normal emission photoemission spectra (after linear background subtraction) of the QW states corresponding to 1 ML and 2 ML of Ag on V(100) recorded at 60 K. Both spectra are fitted with the Fermi liquid lineshape.

Fig. 1 shows the photoemission spectra from 1 and 2 ML films (shown overlaid) taken at 60 K. The lineshape of the peak from the QW state is completely determined by the photo-hole self-energy $\Sigma(\mathbf{k},\omega)$. Depending on the rate of change of the imaginary part, the spectrum may be either Lorentzian-like (Im$\Sigma$ approximately constant) or asymmetrical. If Im$\Sigma$ changes strongly with energy, the lineshape may even acquire a "two-peak" structure [2]. In Fig. 1 the experimental spectra are fitted with the "Fermi liquid" lineshape: $2\text{Im}\Sigma(\omega)=\Gamma_0+2\beta\omega^2$. The energy-independent term, $\Gamma_0$, represents the sum of impurity (or defect) scattering and phonon scattering terms (the phonon contribution being constant at a specific temperature for all energies studied here) and the quadratic term is the electron-electron contribution. Both peaks are fitted by electron-electron coupling parameters that are significantly larger than in bulk-like silver films [7]; $\beta\approx0.03$ to $0.04$ eV$^{-1}$ for 1ML, and $\beta\approx0.08$ eV$^{-1}$ for the 2 ML QW state. These fits lead to values for the constant term $\Gamma_0$ of approximately 150-200 meV for the 1 ML QW state and 50 meV for the 2 ML QW state.

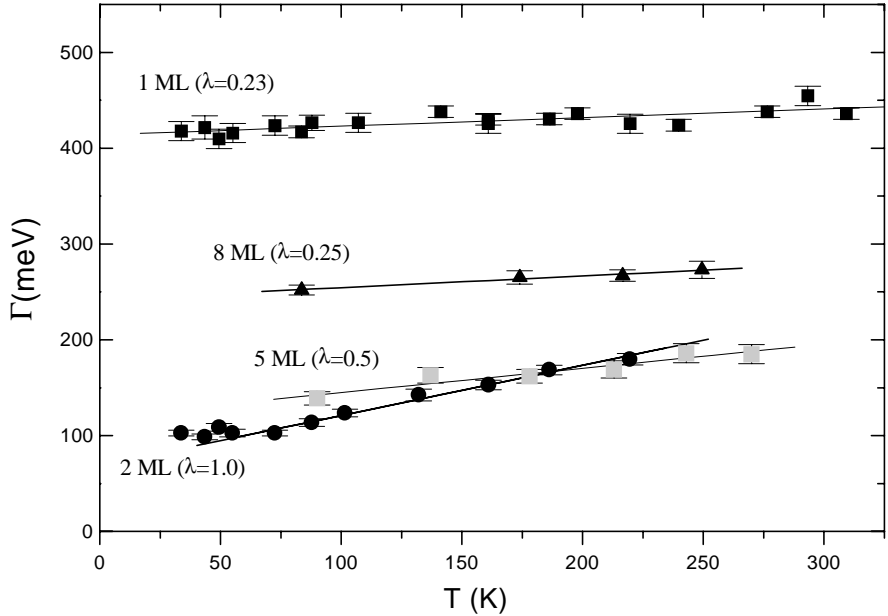

**Fig.2** The peak widths of QW state photoemission peaks plotted as a function of substrate temperature. From the linear fit of each set of data the corresponding electron phonon coupling constant ($\lambda$) was calculated and is shown in brackets. For the sake of clarity not all sets of data are shown.

However, for the present purpose, the electron-phonon scattering is our primary interest. It can be obtained directly from the dependence of the measured ARPES peak widths on temperature; these data are summarised in Fig. 2. The temperature dependence is approximately linear at high temperatures with a slope of $2\pi\lambda k_B$ where $\lambda$ is the electron-phonon coupling constant. The linear fits superimposed in the data in Fig. 2 show that the gradient changes strongly with the film thickness. In Fig. 3 we plot these experimental $\lambda$ values deduced from these linear gradients (filled circles), as a function of the Ag film thickness $d$. The most prominent feature of the $\lambda(d)$ plot is the change in the coupling strength of the QW states to phonons when the thickness of the silver film is increased from 1 ML to 2 ML; the corresponding value of $\lambda$ for the 2 ML Ag film is more than four times larger than that for the 1 ML QW state ($\lambda_{1ML} \approx 0.23$, $\lambda_{2ML} \approx 1.0$) and 3 - 4 times larger than the recently reported value measured for a 19 ML Ag film, believed to yield a value characteristic of bulk Ag, grown on Fe(100) [7]. Apart from the prominent maximum in $\lambda$ at 2 ML, the Ag/V(100) system shows an additional maximum of the coupling constant around 5 ML. To our knowledge, this is the first demonstration of oscillations in the electron-phonon coupling strength with the thickness of the metallic film. The only relevant prior measurements are from 1 ML and 2 ML QW states in the Na/Cu(111) system [9]; these showed a small difference for the two layers, while both QW states showed stronger coupling than the corresponding bulk value.

The λ value for the clean V(100) surface ($\lambda_V \approx 1.45$) is also shown in Fig.3. It is derived from the mass enhancement of the surface state (band dispersion) near the Fermi level in the same manner as in the earlier surface state studies of Mo(110) [3] and Be(0001) [2, 5]. This value represents the strongest coupling measured in ARPES so far and is significantly larger than the bulk value for V of 0.8 [16]. However, this large electron-phonon coupling constant of the vanadium substrate cannot explain the maximum in the coupling constants for the thin film QW states at 2 ML. If the large $\lambda_{2ML}$ value were solely due to the interaction of the silver overlayer with the electronic and phonon system of the underlying vanadium substrate, we would expect the coupling in the 1 ML film to be even stronger. Notice too, that the earlier study of the Na/Cu(111) system found electron-phonon coupling of the QW states for the 1 ML and 2 ML Na films to be found significantly larger than the bulk coupling constants of both Cu and Na, while the coupling constant for the surface state on Cu(111) is essentially identical to that of bulk Cu [4]. These results were taken to imply that the increased electron-phonon coupling strength has its origin in the low dimensionality of the films, rather than in any interaction with the substrate [9].

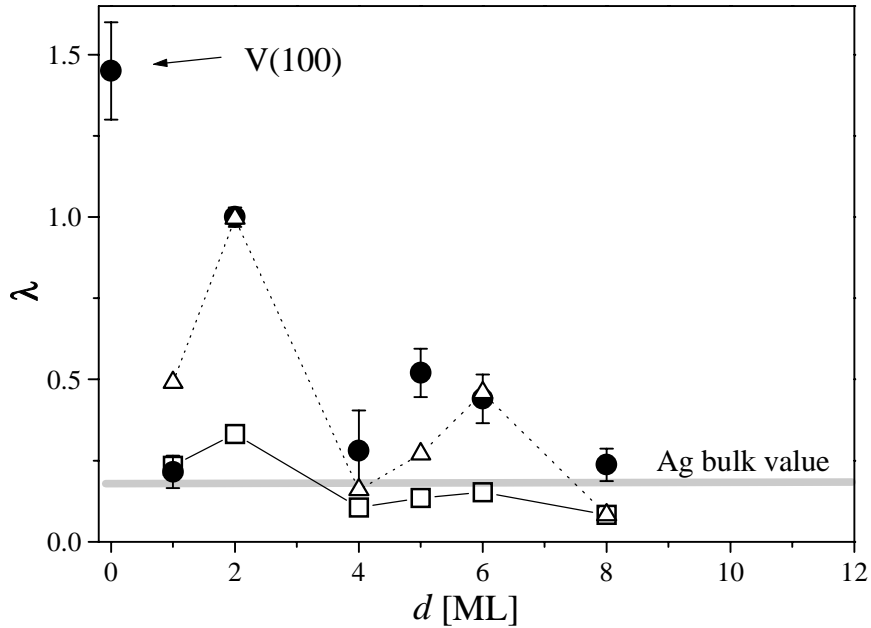

**Fig.3** The electron phonon coupling constant (λ) values obtained from fig. 2 shown as a function of silver film thickness (solid circles). Calculated values of λ assuming an effective mass m*=1 for all QWS are shown as open squares; calculated values including experimentally-determined values of the effective mass for each QW state are shown as open triangles. The experimental value of lambda for bulk vanadium is shown at zero film thickness while the bulk silver value is indicated by the gray horizontal line.

In order to understand the origin of the increased and oscillatory thickness-dependent electron-phonon coupling in our Ag on V(100) QW states, we have performed calculations for a simple model [17] in which the main contribution to the hole lifetime is assumed to be the interaction of the photo-hole with the oscillations of the potential step at the interface between the vacuum and the silver film due to thermal vibrations of the surface atoms. It can be shown that the phonon-induced lifetime $\tau_i$ of the photo-hole created in the QW band $i$, can be written as:

$$\frac{1}{\tau_i} = \frac{m_{x,y}}{M} \frac{A_c}{\hbar^2 |\omega_0|} \left[ |T_{i,i}|^2 (n(\omega_0) + 1) + \sum_{f \prec i} |T_{f,i}|^2 (2n(\omega_0) + 1) \right] \quad (1)$$

Here, $f$ and $i$ are discrete indices associated with the number of nodes that a wave function of the corresponding QW state has in the z-direction. $A_c$ is the area of the surface unit mesh of the silver film, $M$ is atomic mass of silver and $n(\omega_0)$ is Bose- Einstein distribution. Equation (1) is derived assuming that the perpendicular surface vibrational mode is characterised by the Einstein frequency $\omega_0$ and is completely localised at the topmost layer. We have assumed $\omega_0$ to be independent of film thickness and taken the dispersion of the QW states to be isotropic and parabolic parallel to the surface plane with a mean effective electron mass, $m_{x,y}$. The effective transition matrix element, $T_{f,i}$, is given by

$$|T_{f,i}|^2 = V_R^2 [\Psi_f(z=0)\Psi_i(z=0)]^2 \quad (2)$$

where $\Psi_f(z=0)$ and $\Psi_i(z=0)$ are the wave functions of the QW states between which the transition takes place, at the position of the pseudopotential step between the film and the vacuum (z=0). $V_R$ is the height of this step at the film-vacuum interface. Fig. 3 shows values of the electron-phonon coupling constant calculated from the slope of the $\hbar/\tau$ function in the high temperature range. The open squares correspond to values of $\lambda$ calculated assuming that the effective mass of all the QW states around the centre of the Brillouin zone is equal to one, while the open triangles show values of $\lambda$ obtained when the experimentally-measured effective masses for each QW state are used.

These theoretical results show the oscillation of the coupling constant to be present, even if free-electron like dispersion of all the QW states is assumed. In this case the changes in the coupling appear to arise mainly from the different localisation of the states, and specifically the amplitude of the associated wave-functions at the surface barrier. These amplitudes are influenced by the QW state binding energy, the more shallowly-bound states extending further into the vacuum, and by the degree of localisation as determined by the film thickness, thicker films having more extended states which (when normalised) have lower amplitudes at the surface. For example the largest binding energy (approximately 1.5 eV) QW states are those in the 1 ML and 4 ML films which show weak electron-phonon coupling. The lowest binding energy state (0.2 eV) is that in the 6 ML film, but while this shows a higher coupling, the deeper (0.7 eV) but more localised state in the 5 ML film has slightly stronger coupling. The strongest coupling is found for the much more localised state in the 2 ML film with a comparable (0.6 eV) binding energy. There is also some influence on the coupling in this simple theoretical model associated with differences in the available phase space for the photo-hole decay. In particular, in the case of a 1 ML Ag film only a single QW state exists so only intra-band transitions are possible, while for the thicker films, both, inter- and intra-band transitions can occur. However, calculations allowing only intra-band transitions in all films produce results similar to those shown in Fig. 3, indicating that this phase space consideration is secondary.

Of course, the model presented here is very simple. Being based only on a surface-related process, its contribution to the electron-phonon coupling in the QW states vanishes as the number of layers goes to infinity; no account is taken of the interaction with the bulk phonons. Nevertheless, this very simple version of the calculations reproduces the main qualitative features of the experimental data, and thus appears to identify the key underlying physics. In this regard we note that much better quantitative agreement is obtained when the measured effective mass for each QW state is included in the calculations.

In conclusion, we have shown that the electron-phonon coupling constant for silver films of only a few ML thickness oscillates strongly as the thickness is varied. A particularly high value of this constant, $\lambda$ is found for the 2 ML film. The oscillation and the high values of electron-phonon coupling constant found in the silver films deposited on V(100) have been explained in terms of the hole interacting with thermally-induced oscillations of the potential step at the adlayer-vacuum interface, and the variations in coupling of the QW states to this vibrational mode as the degree of localisation (including that determined by the QW state binding energy) varies with film thickness.

The authors acknowledge the support of the British Council and the Ministry of Science of Croatia (ALIS project, grant Zag/984/CRO/042), the Ministry of Science of Croatia (grant 003501108), the Engineering and Physical Sciences Research Council and the U.S. Department of Energy Contract No. DE-AC02-98CH10886.